# Deep Learning Reconstruction for Accelerated Radial CEST MR Fingerprinting

Karandeep Cheema[1], Victor Murray[1], Ricardo Otazo[1,2], Ouri Cohen[1]

1 Department of Medical Physics, Memorial Sloan Kettering Cancer Center, New York, NY, USA

2 Department of Radiology, Memorial Sloan Kettering Cancer Center, New York, NY, USA

**Corresponding author:** Ouri Cohen, PhD, Department of Medical Physics, Memorial Sloan Kettering Cancer Center, 320 East 61st Street, New York, NY 10065, USA; coheno1@mskcc.org

**Short title:** Movienet for accelerated CEST-MRF

**Article type:** Research Article

Main-text word count: 2996

## Abstract

**Purpose:** To accelerate the acquisition of chemical exchange saturation transfer magnetic resonance fingerprinting (CEST-MRF) using temporal deep learning reconstruction.

**Methods:** Ten healthy volunteers underwent repeated (test–retest) golden-angle radial CEST-MRF acquisitions. Each acquisition used a schedule of 30 measurements, and 34 spokes were used per measurement. The Movienet reconstruction network, originally developed for motion-resolved imaging, was adapted to reconstruct accelerated golden-angle radial CEST-MRF data. To test different acceleration factors, separate Movienet models were trained using 1, 2, 4, 8, 16, 20, and 34 spokes per measurement. Movienet exploited temporal and coil correlations by concatenating the time and coil dimensions in the input. For comparison, a GRASP-based locally low-rank (GRASP-LLR) reconstruction was performed. Quantitative DRONE fitting was applied to both Movienet and GRASP-LLR reconstructions to obtain tissue parameter maps. Movienet-8 was further evaluated using leave-one-subject-out (LOSO) analysis. Prospectively acquired 8-spoke datasets were evaluated alongside paired 34-spoke baseline acquisitions, and Movienet-8 and Movienet-34 were additionally compared with a 64-spoke acquisition.

**Results:** In the primary held-out retrospective series, median Pearson correlation between the Movienet and GRASP-LLR reconstructions was 0.795, 0.947, 0.974, 0.977, and 0.981 for 1, 8, 16, 20, and 34 spokes, respectively; corresponding NRMSE values were 14.7%, 8.2%, 5.9%, 5.3%, and 5.1%. Across 10 LOSO volunteers, median subject-level image correlation was 0.965 and NRMSE was 7.5%. Movienet inference for a 30-slice series required 1.7 s, whereas GRASP-LLR reconstruction required approximately 45 min.

**Conclusion:** Movienet enables whole-brain radial CEST-MRF acquisition in 3 minutes, with network inference for image reconstruction requiring less than 2 seconds.

## 1. Introduction

Chemical exchange saturation transfer (CEST) MRI detects labile solute protons indirectly through exchange-mediated changes in the bulk-water signal [1], [2]. The resulting contrast is sensitive to molecular concentration, exchange rate, relaxation, and tissue microenvironment [1], [2]. Conventional CEST methods remain constrained by acquisition duration, multistep processing, and semiquantitative measures that depend on acquisition conditions. These limitations can be addressed by combining CEST with MR fingerprinting (MRF) in which acquisition parameters are varied over time and the resulting contrast-varying signal is used to quantify the underlying tissue properties [3], [4]. CEST-MRF quantification was established using EPI k-space acquisition and dictionary matching [5] and adapted using deep learning matching [6]. Deep learning matching estimates tissue parameters directly from the measured fingerprints, avoiding an exhaustive search through a large multidimensional dictionary and reducing the computational and memory demands of quantification [6]. Related work investigated acquisition-schedule design [7], deep-learning quantification of treatment response [8], and acquisition-schedule optimization [9]. Other studies demonstrated multislice whole-brain imaging [10], developed deep-learning methods for CEST and MT quantification [11], and detailed the deep-MRF workflow [12].

Although EPI provides fast k-space sampling, it is also sensitive to susceptibility differences that can lead to geometric distortions in the resulting images. In contrast, golden-angle radial sampling is advantageous because the center of k-space is repeatedly sampled and the readout is short, which renders it less susceptible to geometric distortion than EPI [13]. The main drawback of radial acquisitions is that sampling is considerably slower than EPI since multiple spokes must be acquired for each image. Iterative radial reconstruction also requires coil compression, sensitivity estimation, non-Cartesian encoding, density compensation, and temporal regularization before quantitative fitting.

A radial CEST-MRF pulse sequence was recently introduced [14] to provide whole-brain quantitative CEST and MT maps using 34 spokes per measurement. The 34-spoke whole-brain acquisition used in the present study required approximately 11 minutes. The radial data were reconstructed with an iterative GRASP-based locally low-rank (LLR) reconstruction [14], [15], which required approximately 45 min using a GPU-accelerated implementation. The radial acquisition yielded undistorted tissue maps, but the relatively long scan time is a significant limitation for clinical adoption and of patient throughput. Reducing the number of spokes per measurement with GRASP-LLR reduces scan time but increases blurring or streaking, which is a common effect in compressed sensing techniques.

The application of deep learning for reconstruction of undersampled radial k-space has been demonstrated to overcome the limitations of compressed sensing techniques, such as GRASP-LLR reconstruction. For example, Movienet was demonstrated to improve the performance of compressed sensing for motion-resolved dynamic MRI using golden-angle radial k-space acquisition [16]. However, CEST-MRF presents a different and more challenging reconstruction problem than smooth changes due to respiratory motion, because the acquisition parameter variations in the schedule produce discontinuous contrast changes to encode tissue properties.

The aim of this work is to significantly accelerate golden-angle radial CEST-MRF acquisition by adapting the Movienet architecture. Feasibility of the proposed approach is evaluated using CEST-MRF acquisitions in healthy subjects with retrospective and prospective accelerations. Validation is performed based on quantitative tissue maps in comparison to compressed sensing reconstruction.

## 2. Methods

### 2.1. Radial CEST-MRF pulse sequence

The original radial CEST-MRF pulse sequence [14] was utilized for this work. In each measurement of the acquisition, a Gaussian pulse train with frequency set to the amide resonance frequency (3.5 ppm) is used to saturate the exchangeable amide protons. A slice-selective RF excitation pulse set to the water frequency (0 ppm) is used to tip the magnetization to the transverse plane. The resulting signal is read using a stack-of-stars radial k-space sampling scheme with the same spoke angle used for all 30 slices. The schedule of saturation powers, durations, RF excitation flip angles and TRs used for each of the 30 measurements was defined via schedule optimization [9]. Following acquisition of all 30 measurements, the spoke angle was updated by integer multiples of the golden-angle to ensure consecutive spokes in each measurement which also enabled retrospective undersampling of the data by including or discarding the data from individual spokes.

### 2.2. Healthy human subject studies

Ten healthy volunteers (1 male, 9 females; mean age, 28.0 ± 3.1 years) were scanned using repeated 34 spokes. The study was approved by the Memorial Sloan Kettering Cancer Center Institutional Review Board, and all participants provided written informed consent. Each volunteer was scanned twice (test–retest) using 34 spokes per measurement. Six of these same volunteers underwent additional prospective 8- and 16-spoke acquisitions; the 8-spoke scans and paired first (i.e., test) 34-spoke acquisitions were evaluated here. A separate 64-spoke healthy-volunteer examination, excluded from training, validation, and leave-one-subject-out (LOSO) evaluation, provided the higher-density reference. Each 34-spoke acquisition was retrospectively undersampled to 1, 2, 4, 8, 16, 20, and 34 (i.e., no undersampling) spokes per measurement.

## 2.3. MRI experiments

All experiments were conducted on a Signa Premier 3T scanner (GE Healthcare, Waukesha, WI) with body-coil transmission and a 48-channel head receiver coil. The following acquisition parameters were used: field of view was 280 × 280 $mm^2$ with a 256 × 256 matrix, approximately 1.1 × 1.1 $mm^2$ in-plane resolution, 3-mm slice thickness, 30 slices, echo time of 2.65 ms, and receiver bandwidth of 62.5 kHz.

## 2.4. Training, validation, and testing datasets

A testing dataset was defined using the 34-spoke test and retest acquisitions from one subject. The data from the remaining nine subjects (9 subjects × 2 scans/subject × 30 slices/scan = 540 slices) were randomly split into 432 training slices and 108 validation slices (80/20 split).

### 2.4.1 Leave-one-subject-out (LOSO) experiment

The generalizability of Movienet-8 to unseen subjects was evaluated using ten LOSO folds. Each fold used eight subjects for training, one for validation, and one for testing, with both acquisitions from each volunteer retained in the same partition.

### 2.4.2 Prospectively acquired data analysis

Prospectively acquired 8-spoke data and paired baseline 34-spoke acquisitions were analyzed. Although the GRASP-LLR reconstruction served as a reference, it is also susceptible to aliasing artifacts for insufficient data. Therefore, the 8 and 34-spoke reconstructions were also compared to reconstruction of a 64-spoke acquisition.

## 2.5. GRASP-LLR reference

### 2.5.1 Image reconstruction

GRASP-LLR reconstruction was used as the reference for network training. The measured data were reconstructed offline with a GRASP-based LLR formulation [14], [15]. The 48-channel data were compressed to 8 virtual coils by principal-component coil compression [17]. Coil sensitivities were estimated using reference images obtained by coil-by-coil nonuniform fast Fourier transform (NUFFT) [18] reconstruction of all spokes followed by adaptive array combination [19]. For each slice, the reference series was obtained by solving Equation (1) [14], with nuclear-norm regularization of local space-time blocks [20].

$$m_{ref} = argmin_m ||Em - d||_2^2 + \lambda \Sigma_i \|R_i m\|_* \quad (1)$$

In Equation (1), $m$ is the image-series optimization variable and $m_{\mathrm{ref}}$ is the reconstructed reference image series, $d$ is the acquired radial k-space data, $E$= FC is the encoding operator containing the nonuniform Fourier operator F and coil-sensitivity operator C, $R_i$ extracts the local space-time block at location i, λ is the regularization weight, and $\|\cdot\|_*$ denotes the nuclear norm or sum of singular values.

### 2.5.2 Image post-processing

Image postprocessing was performed in MATLAB R2025b (MathWorks, Natick, MA, USA). Each 34-spoke GRASP-LLR frame was denoised with MATLAB's pretrained DnCNN implementation, the series was jointly scaled to [0,1], and image sharpening was applied. The post-processed GRASP-LLR series served only as the supervised training target. Original GRASP-LLR reconstructions, without denoising or sharpening, were used as evaluation references.

### 2.5.3 Quantitative map generation

The unprocessed 34-spoke GRASP-LLR reconstructions (i.e., without denoising or image sharpening) were supplied to a DRONE [21] network trained for CEST-MRF parameter estimation for reference quantitative-map fitting. The network was trained as described in Ref. [14]. The output parameters were water T1 (T1w), water T2 (T2w), amide exchange rate (ksw), semisolid exchange rate (kssw), amide pool fraction (fs), and semisolid pool fraction (fss).

## 2.6. Movienet training and validation

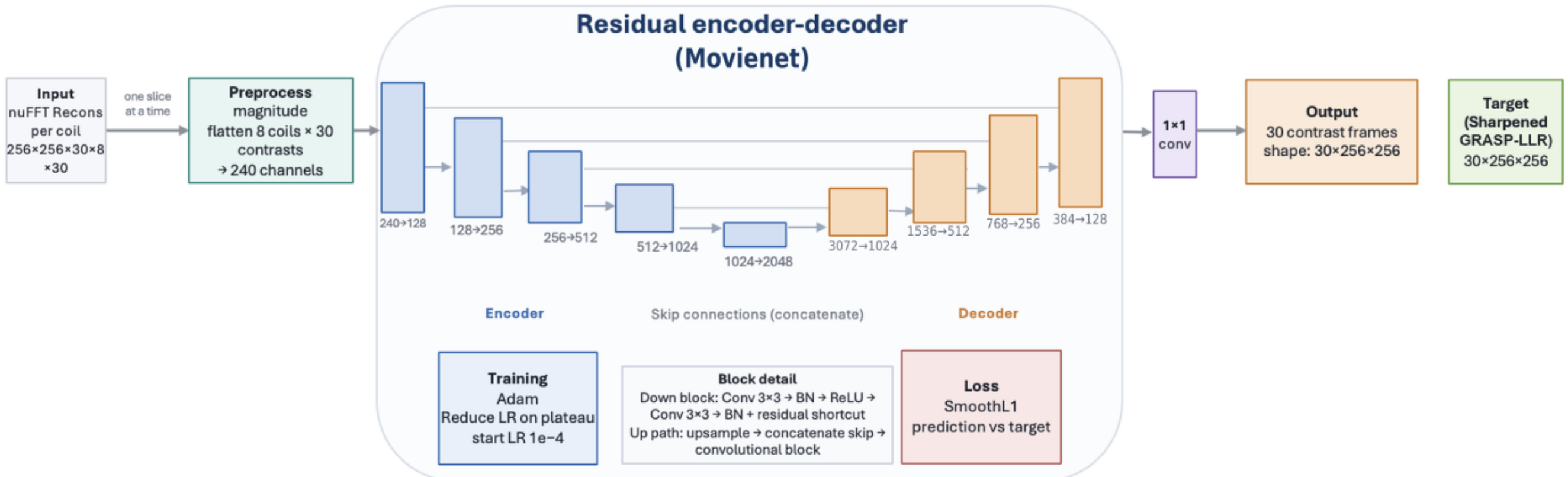

**Figure 1.** Movienet architecture and supervised reconstruction pathway. For each slice, undersampled magnitude images from 8 virtual coils and 30 measurements are flattened to 240 channels and processed by a residual encoder-decoder. The output is a 30-frame image series trained against the matched sharpened GRASP-LLR target with Smooth L1 loss.

Figure 1 shows the Movienet residual encoder-decoder [16]. For CEST-MRF, the input was reshaped to 240 channels (8 virtual coils × 30 measurements) and the output to the corresponding 30-frame image series. A separate model was trained for each spoke count using the Adam optimizer [22], a learning rate of $10^{-4}$, batch size 4, Smooth L1 loss, and 2000 epochs. Each spoke-specific model was trained on a single NVIDIA RTX A5500 GPU (24 GB) in a server equipped with eight such GPUs, and the checkpoint with the lowest validation loss was retained.

## 2.7. Evaluation of image quality and tissue-map agreement

The Movienet outputs were compared with matched GRASP-LLR references. A custom intensity-based mask was used to isolate the brain region for evaluation. For each pixel in an axial slice, the 80th percentile of its intensities across the 30 reference frames was calculated to form a single anatomical proxy image. This image was normalized and smoothed, and Otsu thresholding [23], constrained to a normalized threshold of 0.08–0.16, separated the foreground from background. Hole filling, largest-component selection, morphological opening and closing, and convex-hull filling produced the brain-region mask. An 8-pixel in-plane erosion reduced peripheral scalp/skull and partial-volume contributions. Each comparison used the same reference-derived mask for the reference and reconstructed images.

Image agreement was evaluated using structural similarity index (SSIM) [24], peak signal-to-noise ratio (PSNR), root-mean-square error (RMSE), normalized RMSE (NRMSE), mean absolute error (MAE), signed bias, and Pearson correlation. NRMSE used inter-percentile-range normalization [25] and was defined as 100 × RMSE / (P99 − P1), where P99 and P1 are the 99th and 1st percentiles of the reference values within the mask for the corresponding slice and frame or parameter map. This definition normalizes RMSE to the reference percentile range rather than the reference mean.

Quantitative CEST/MT maps were computed from the Movienet series using the same DRONE network as in Section 2.5.3 [21]. Map agreement was assessed using difference maps, SSIM, PSNR, MAE, NRMSE, and Pearson correlation. For aggregate MAE, each tissue parameter was normalized by the corresponding tissue parameter range. Quantitative maps were displayed with fixed parameter-specific color limits across methods and slices. Signed difference maps used fixed symmetric limits centered on zero. Pool fractions and their difference maps were expressed as percentages.

## 3. Results

### 3.1. Effect of spokes per measurement

As expected, increasing the number of spokes per measurement increased the image quality metrics (Figure 2). The median Pearson correlation between the reconstructed and reference images increased from 0.795 at 1 spoke to 0.947 at 8 spokes, 0.974 at 16 spokes, 0.977 at 20 spokes, and 0.981 at 34 spokes. Corresponding NRMSE values were 14.7%, 8.2%, 5.9%, 5.3%, and 5.1%.

Across the ten LOSO test subjects, median subject-level image correlation was 0.965 (range, 0.938–0.977), and median NRMSE was 7.5% (range, 6.5%–11.7%). The individual subject-level results are reported in Supporting Table S1.

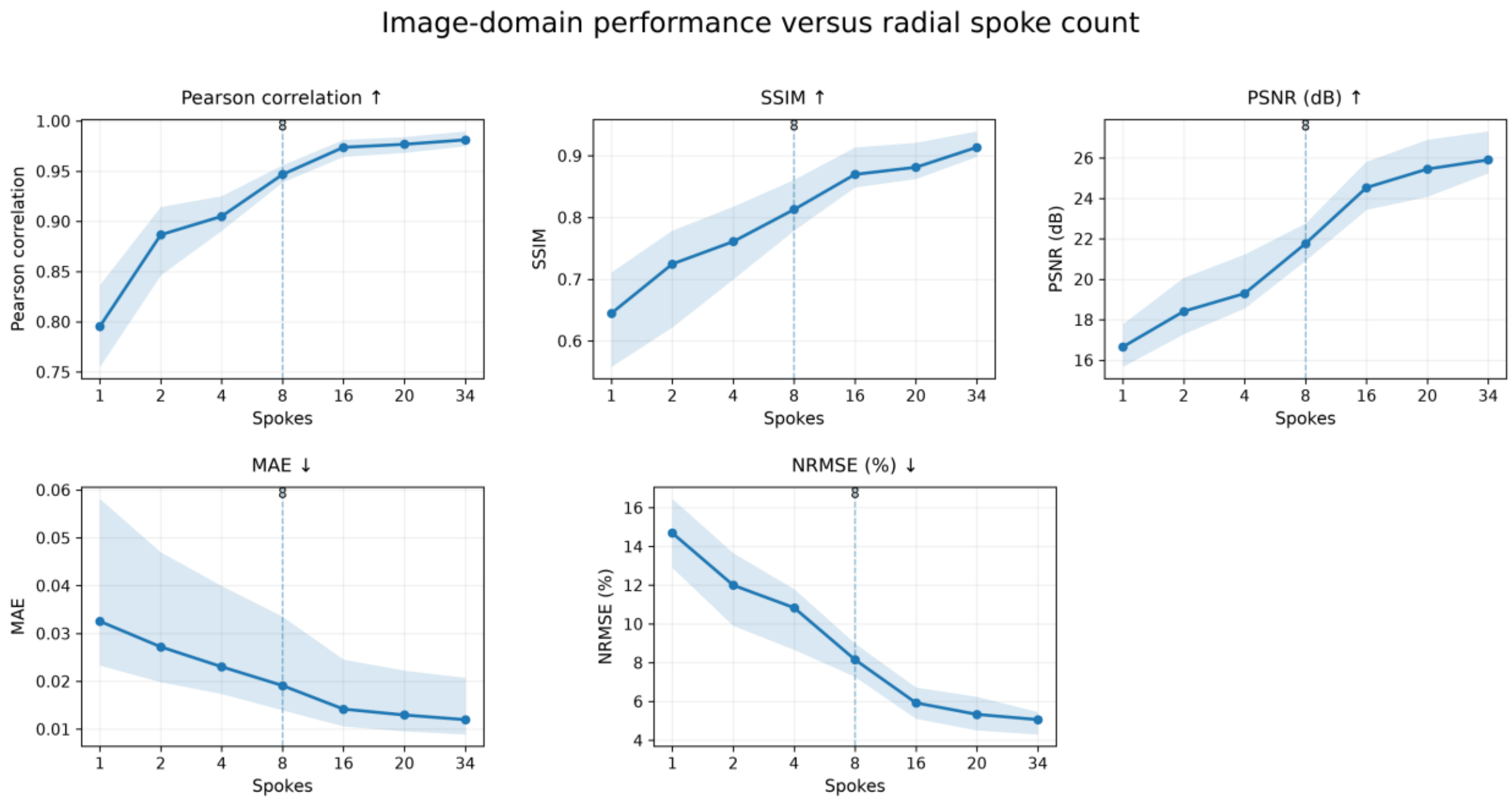

**Figure 2.** Image-domain performance versus radial spoke count in the primary held-out retrospective examination. Points and lines show the median across 900 matched slice-frame pairs (30 slices × 30 frames), and the light-blue band shows the interquartile range. NRMSE is displayed as a percentage. The vertical dashed line marks the 8-spoke operating point selected for prospective acquisition. Arrows in the panel titles indicate whether increasing or decreasing values represent improved performance for each metric.

### 3.2. Image reconstruction of retrospective data

Aggregate image metrics are reported in Table 1. Representative retrospective reconstructions at inferior, middle, and superior brain levels along with the GRASP-LLR reference are shown in Figure 3. Representative retrospective reconstructions using Movienet-16 and Movienet-20 are shown in Figure S1. Differences between the Movienet output and the reference were mostly concentrated in regions with signal voids (e.g., the skull–brain boundary), where sharpening of the training targets may contribute to the differences. At 8 spokes, Movienet-8 showed less conspicuous streaking and blurring than the matched GRASP-LLR reconstruction (Figure 3). Both the 8-spoke and 34-spoke reconstructions showed strong visual agreement with the reference images.

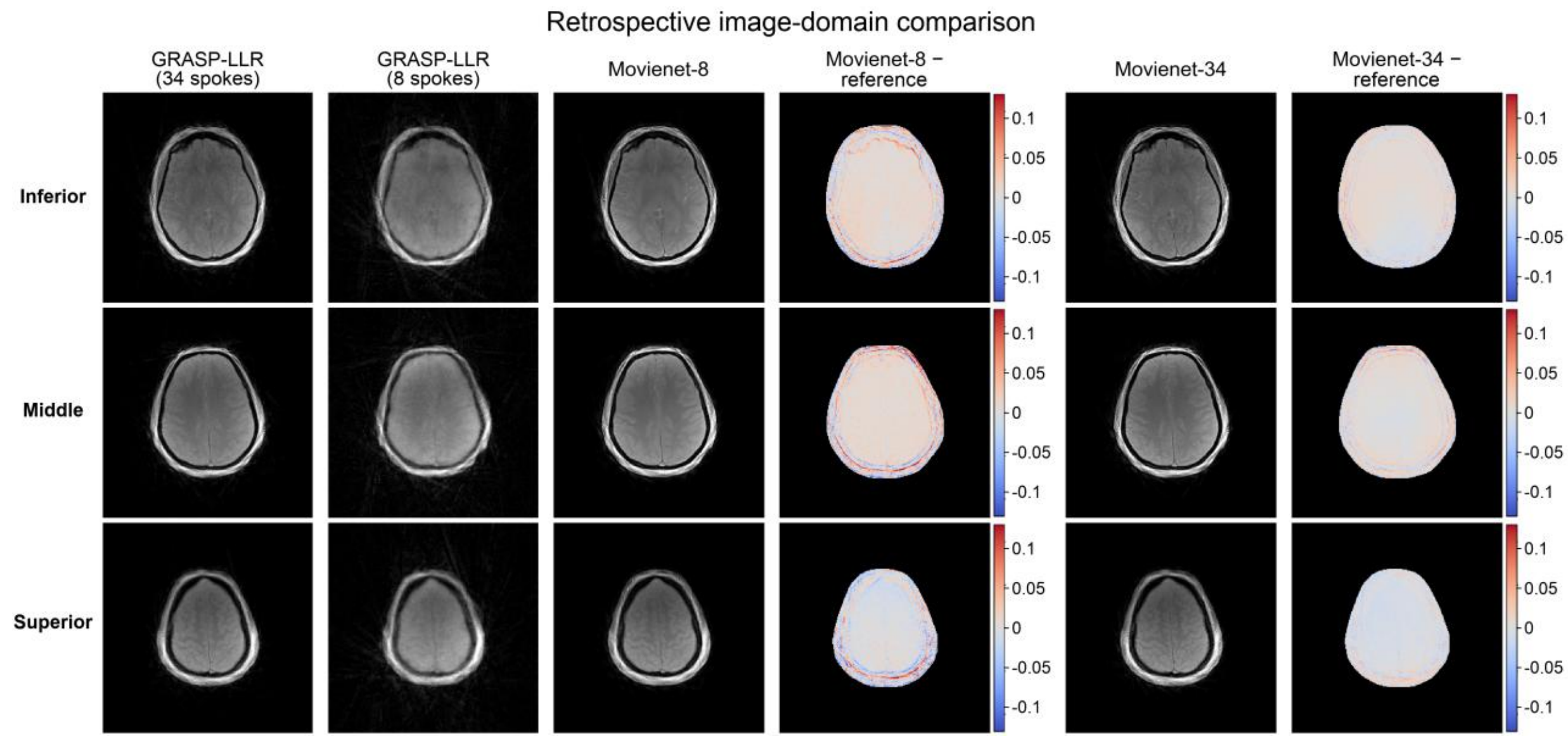

**Figure 3.** Retrospective image-domain comparison at frame 20 in representative inferior, middle, and superior axial slices from the held-out healthy test volunteer. The 34-spoke GRASP-LLR reference images are compared with 8-spoke GRASP-LLR, Movienet-8, and Movienet-34 reconstructions; difference images are shown for both Movienet reconstructions.

### 3.3. Quantitative maps of retrospectively sampled data

Movienet-derived quantitative maps preserved the spatial distributions of the tissue parameters in the held-out healthy examination (Figure 4). Both 8- and 34-spoke reconstructions retained parenchymal and ventricular contrast. Residual streaking visible in the reference maps was less conspicuous in the Movienet outputs, producing spatially nonuniform differences. Movienet-34 yielded lower aggregate map NRMSE than Movienet-8 (Table 2).

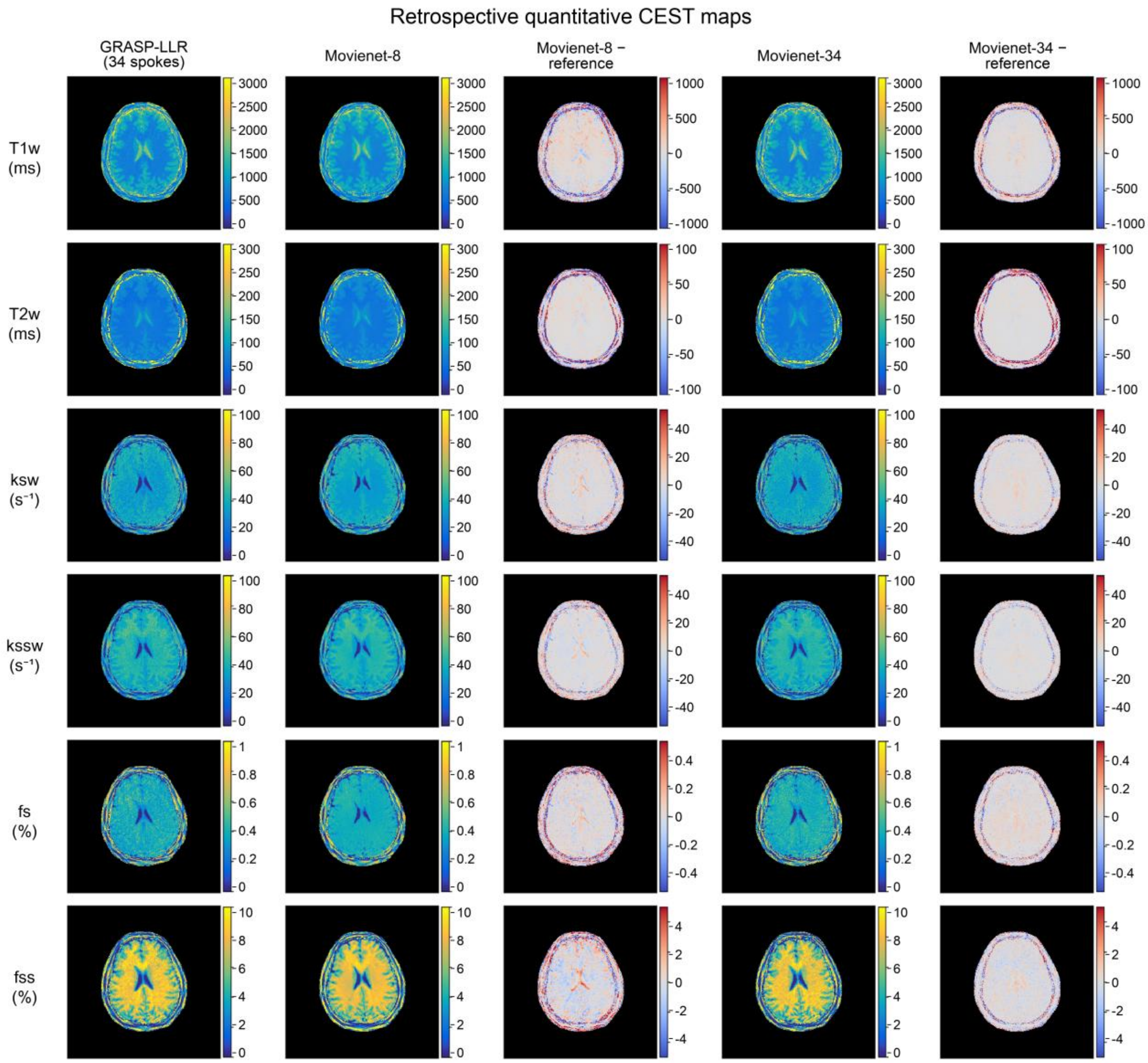

**Figure 4.** Retrospective quantitative CEST map-domain comparison in the held-out healthy test volunteer. Reference maps fitted from original, unsharpened 34-spoke GRASP-LLR images are compared with Movienet-8 and Movienet-34 maps for T1w, T2w, ksw, kssw, fs, and fss.

### 3.4. Evaluation of prospectively acquired data and tissue maps

Prospective 8-spoke acquisitions were evaluated in six volunteers: five who contributed training and validation slices and the held-out test volunteer. Table 1 reports Movienet-8 image agreement with the same-acquisition 8-spoke GRASP-LLR reconstruction. Table 2 reports map agreement for prospective Movienet-8 and paired baseline Movienet-34 against each volunteer's first (i.e., test) 34-spoke GRASP-LLR acquisition. The latter comparison includes acquisition differences as well as reconstruction differences. Representative Movienet-8 reconstructions and tissue maps are shown in Figures 5 and 6.

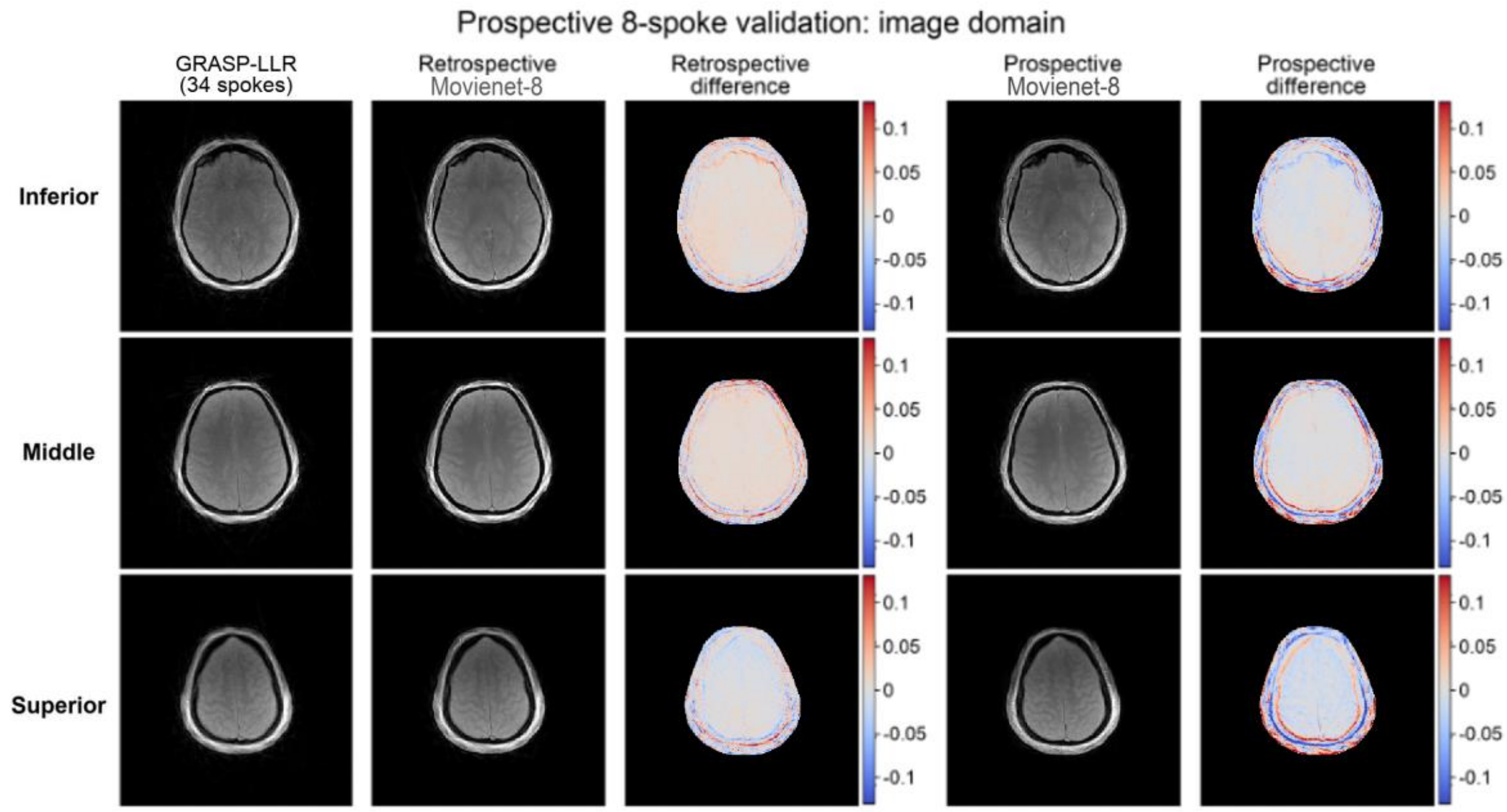

**Figure 5.** Evaluation of the prospectively acquired 8-spoke data in 3 representative slices of a healthy subject. Rows show representative inferior, middle, and superior axial slices for one frame of the MRF schedule. From left to right, columns show the GRASP-LLR reference reconstruction with 34 spokes, retrospective Movienet-8 reconstruction, retrospective difference image, prospective Movienet-8 reconstruction, and prospective difference image.

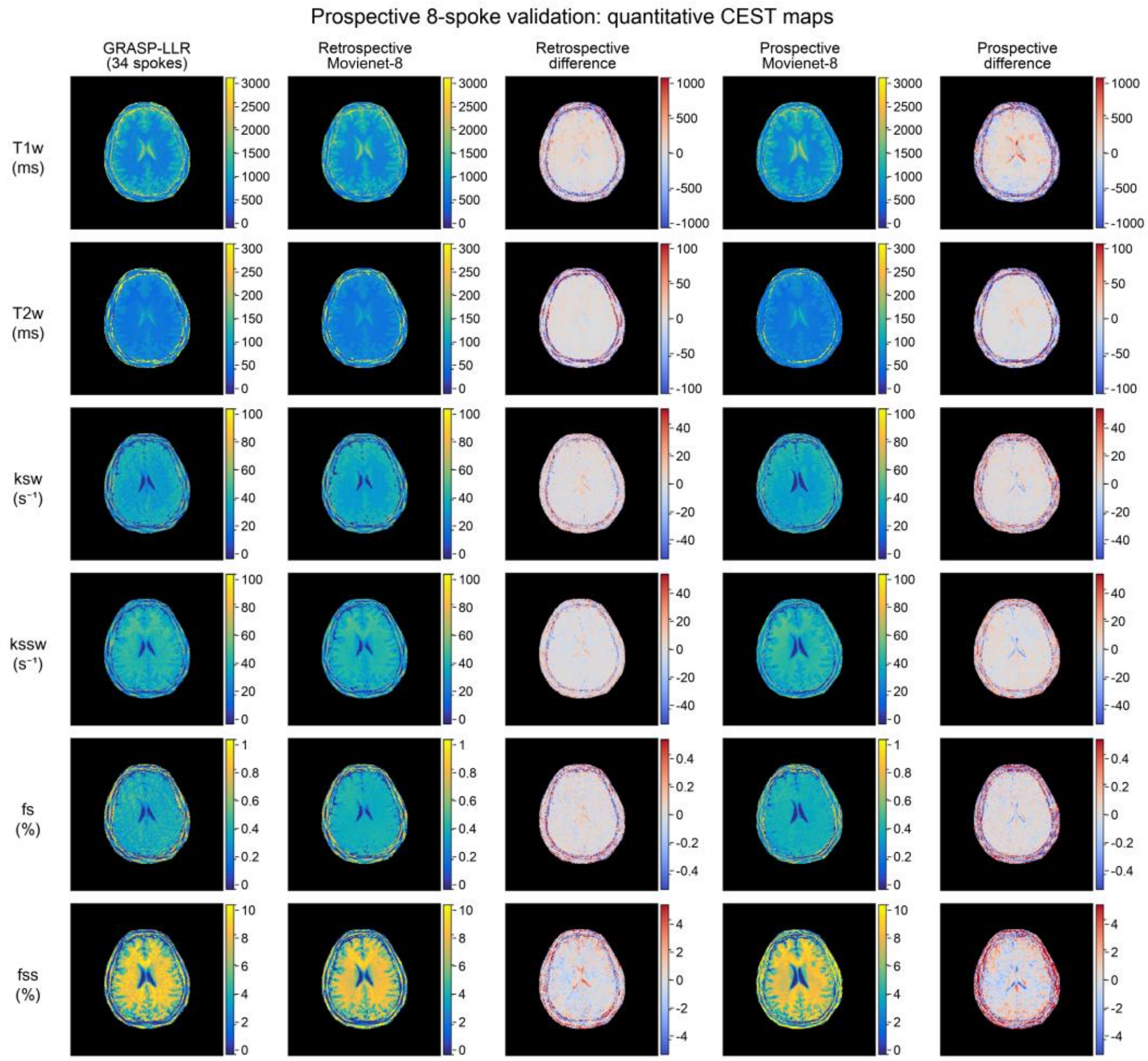

**Figure 6.** Quantitative CEST-MRF maps from a prospective 8-spoke acquisition in a representative slice. From left to right, columns show the reference map fitted from the original paired 34-spoke GRASP-LLR acquisition, retrospective Movienet-8 map, retrospective difference map, prospective Movienet-8 map, and prospective difference map. Rows show T1w, T2w, ksw, kssw, fs, and fss. Parameter and difference-map units and color scales are displayed beside the panels.

### 3.5. Evaluation of Movienet reconstruction in comparison to 64-spoke acquisition

Both Movienet-8 and Movienet-34 preserved anatomy across the selected slices, while Movienet-34 produced smaller differences from the reference (Figure 7). The quantitative maps retained broad anatomical distributions, with spatially varying differences from the 64-spoke reference (Figure 8).

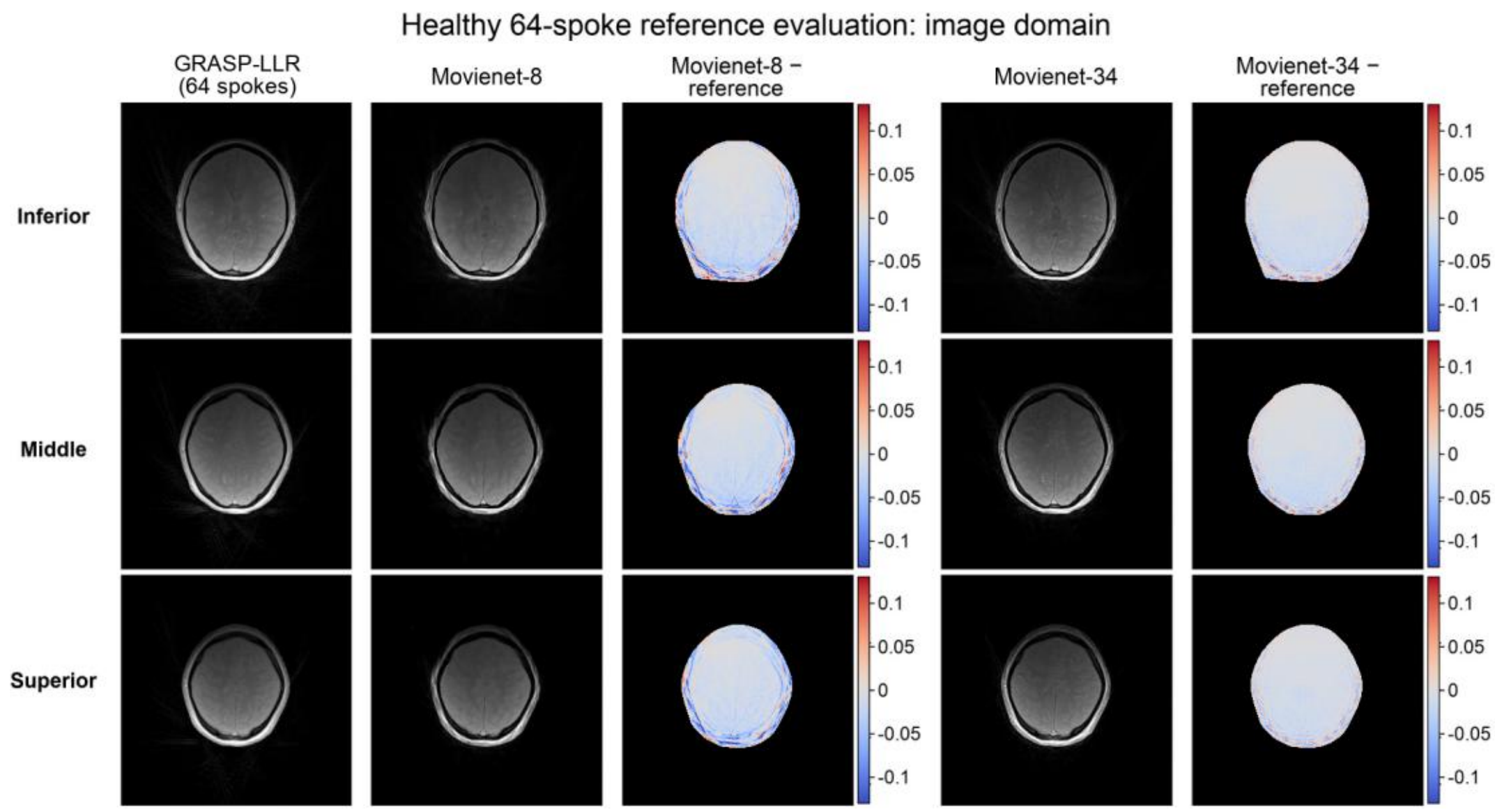

**Figure 7.** Healthy 64-spoke high-sampling-density evaluation. Image-domain comparison at frame 20 across three representative axial slices, using the 64-spoke GRASP-LLR reconstruction as the reference.

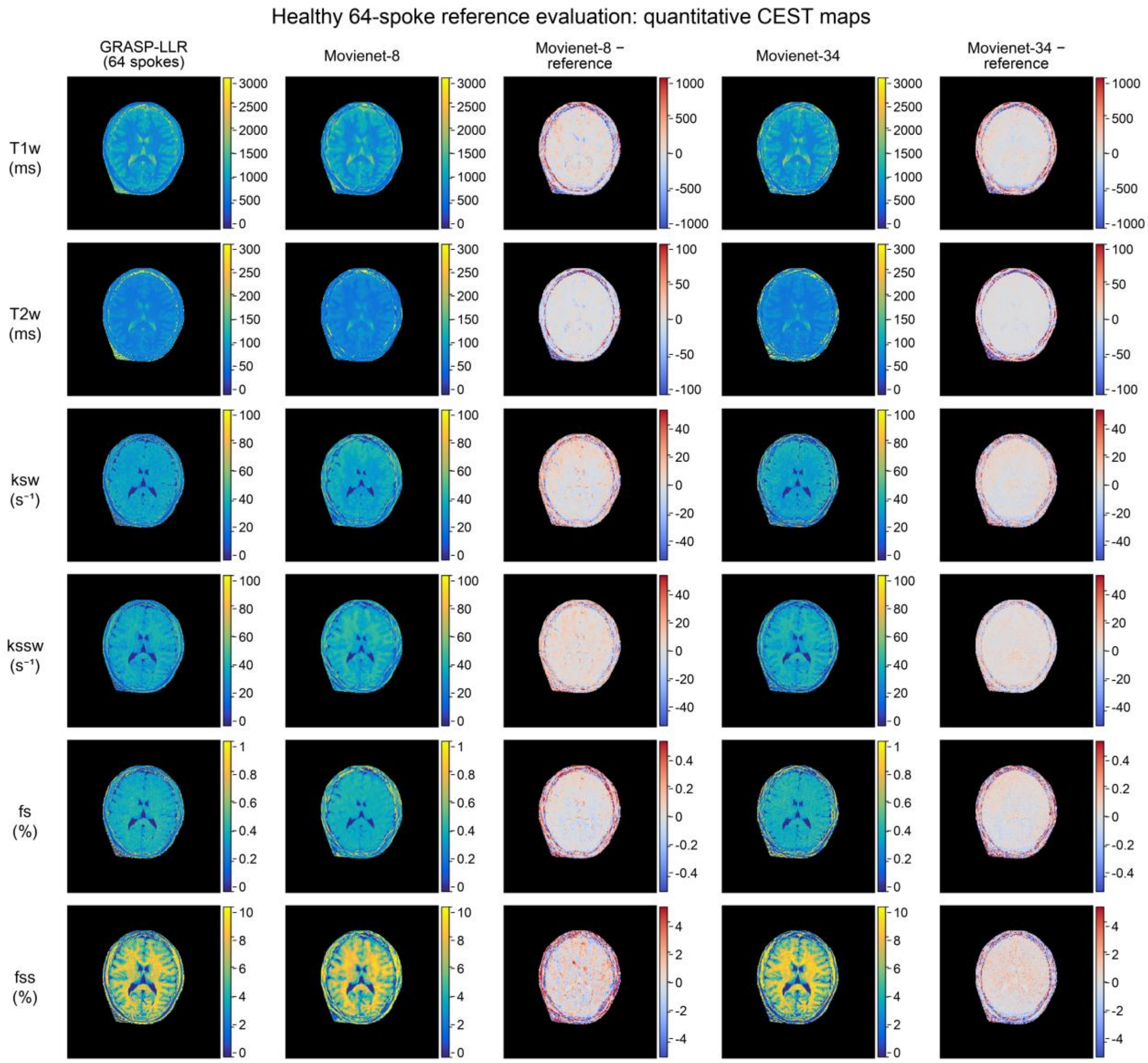

**Figure 8.** Quantitative CEST map-domain comparison in a representative slice; the reference maps were fitted from the 64-spoke GRASP-LLR reconstruction and compared with Movienet-8 and Movienet-34 maps.

### 3.6. Image quality metrics

Aggregate image and tissue map metrics are reported in Tables 1 and 2. In the retrospective image evaluation, Movienet-8 yielded mean Pearson correlation of 0.953 ± 0.026 and NRMSE of 7.1% ± 1.8%; Movienet-34 yielded 0.982 ± 0.011 and 4.6% ± 1.2%, respectively. Movienet-34 also yielded lower aggregate map NRMSE in the retrospective, paired prospective, and 64-spoke evaluations. Tissue map agreement was lower than image agreement: retrospective map correlations were 0.821 ± 0.183 for Movienet-8 and 0.892 ± 0.146 for Movienet-34.

## 4. Discussion

Movienet enables large accelerations of both acquisition and reconstruction times while preserving the temporal signal dynamics that are important for quantitative tissue mapping. Reducing radial sampling from 34 to 8 spokes shortens the acquisition from approximately 11 to 3 min, making the scan duration more practical for clinical imaging. Network inference for the 30-slice series required 1.7 s, compared with nearly 45 min for a GPU-accelerated iterative GRASP-LLR reconstruction. The approximately fourfold reduction in scan time was accompanied by an increase in mean retrospective NRMSE of 2.5 percentage points for image reconstruction (Table 1) and 5 percentage points for quantitative tissue maps (Table 2). Reduced streaking and blurring relative to 8-spoke GRASP-LLR support the accelerated reconstruction approach, although differences from the 34-spoke reference remain (Figure 3). By providing fast reconstructions of undersampled data, the proposed method addresses two key barriers to clinical adoption: scan and reconstruction time.

Reducing scan time can also be achieved by reducing the number of measurements (i.e., the length of each fingerprint) rather than the number of radial readouts in each image. Here, we chose to retain all 30 CEST-MRF measurements to allow Movienet to use the complete sequence of dynamic contrast-dependent images. Reducing the number of spokes increases spatial undersampling within each image while retaining all contrast encodings used for tissue-parameter estimation. Reducing the number of measurements removes contrast encodings from the fingerprint and changes the information available for distinguishing tissue parameters. Since the acquisition parameter schedule has already been optimized for the defined number of measurements [14], and the aim was to obtain quantitative tissue maps, the network was evaluated on its ability to recover the image series from fewer radial samples rather than fewer measurements. Recovery of the complete contrast-dependent image series and the principal tissue-map distributions supports the feasibility of this approach.

The use of prospectively acquired data demonstrates the utility of the model for actual low-spoke acquisitions, extending the evaluation beyond retrospectively undersampled data. Although both retrospective and prospective acquisitions used golden-angle ordering, different spoke angles result from each approach. Golden-angle sampling nevertheless provides approximately uniform angular coverage of k-space. The qualitative similarities between the retrospective and prospective reconstructions (Figures 3–6) support the applicability of the learned reconstruction beyond the specific sampling pattern used. Unlike retrospectively undersampled data, where the input and reference originate from the same acquisition, the prospective map comparisons used separately acquired 8-spoke data and a paired 34-spoke reference. Subject motion and positional differences between these scans can contribute to higher voxelwise differences, particularly at tissue boundaries.

Since a gold-standard reference was not available in vivo, this study used GRASP-LLR reconstruction with 34 spokes per measurement as the reference. However, GRASP-LLR reconstructions are also susceptible to undersampling artifacts arising from limited data, which can affect the resulting image and map metrics. To provide a more densely sampled reference, a dataset acquired with 64 spokes in one healthy volunteer was also evaluated. The results (Figures 7 and 8) show that Movienet-8 retained anatomical features and broad spatial distributions of the quantitative maps, including recognizable gray–white matter contrast and ventricular anatomy, particularly in the $T_1$w and fss maps (Figure 8). Unsurprisingly, Movienet-34 retained finer spatial detail and produced smaller differences from the reference. Mean map NRMSE was 21.4% for Movienet-8 and 16.1% for Movienet-34, a difference of 5.3 percentage points (Table 2).

In previous studies, Movienet was used to reconstruct smoothly varying dynamic signals caused by respiratory motion or contrast enhancement [16], [26], [27]. Reconstruction of CEST-MRF data presents an additional challenge because the contrasts are deliberately optimized for parameter encoding, leading to abrupt contrast changes across measurements. Although the contrasts change, the measurements retain shared anatomical structure and signal dependencies governed by the underlying tissue properties. Movienet processes all 30 measurements jointly, allowing it to exploit these temporal correlations to recover information from undersampled images. The observed recovery of dynamic signal patterns and quantitative tissue maps at 8 spokes per measurement supports the presence of sufficient temporal correlation in CEST-MRF to achieve the acceleration demonstrated here.

Unlike the EPI readouts used in prior CEST-MRF studies [6], [10], [12], the radial CEST-MRF sequence is less susceptible to geometric distortion. However, a drawback of radial imaging is the increased scan time required to avoid streaking artifacts. The proposed reconstruction method reduces this scan-time penalty by recovering the image series from substantially fewer radial spokes. The EPI-based Q-MAJOR-CEST approach achieved whole-brain coverage in approximately 6 min using a multislab acquisition with 3-mm slices [10]. The approximately 3-min acquisition demonstrated here places accelerated radial CEST-MRF in a comparable scan-time range while retaining the radial readout's reduced susceptibility to geometric distortion. Differences in acquisition schedules and imaging protocols should be considered when comparing these scan times.

The principal limitation of this study is the small, single-site cohort and the restriction of training and evaluation to healthy anatomy. Tumors, lesions, or other abnormalities that differ from the training distribution may incur higher error or lead to network hallucinations, so patient studies will be required before diagnostic use can be considered. Incorporating a wider range of anatomy and pathology into training can help address this limitation and is the focus of future work. Alternatively, reductions in scan acceleration or data augmentation with synthetic CEST data, as recently demonstrated [28], [29], offer further strategies for improving reconstruction robustness.

## 5. Conclusions

The proposed novel deep learning reconstruction approach using the Movienet network for golden-angle radial CEST-MRF data enables whole-brain quantitative mapping with a scan time of 3 minutes and less than 2 seconds of network inference for image reconstruction. By significantly reducing both acquisition and reconstruction times, Movienet would enable the integration of golden-angle radial CEST-MRF in the clinical workflow with reduced geometric distortions compared to EPI.

## 6. Acknowledgments

This work was supported by Grant/Award Numbers: R37-CA262662. This manuscript is the result of funding in whole or in part by the National Institutes of Health (NIH). It is subject to the NIH Public Access Policy. Through acceptance of this federal funding, NIH has been given a right to make this manuscript publicly available in PubMed Central upon the Official Date of Publication, as defined by NIH.

## References

[1] M. Zaiss and P. Bachert, “Chemical exchange saturation transfer (CEST) and MR Z-spectroscopy in vivo: a review of theoretical approaches and methods,” Phys. Med. Biol., vol. 58, no. 22, pp. R221–R269, 2013, doi: 10.1088/0031-9155/58/22/R221.

[2] K. M. Jones, A. C. Pollard, and M. D. Pagel, “Clinical applications of chemical exchange saturation transfer (CEST) MRI,” J. Magn. Reson. Imaging, vol. 47, no. 1, pp. 11–27, 2018, doi: 10.1002/jmri.25838.

[3] D. Ma et al., “Magnetic resonance fingerprinting,” Nature, vol. 495, no. 7440, pp. 187–192, 2013, doi: 10.1038/nature11971.

[4] M. E. Poorman et al., “Magnetic resonance fingerprinting Part 1: Potential uses, current challenges, and recommendations,” J. Magn. Reson. Imaging, vol. 51, no. 3, pp. 675–692, 2020, doi: 10.1002/jmri.26836.

[5] O. Cohen, S. Huang, M. T. McMahon, M. S. Rosen, and C. T. Farrar, “Rapid and quantitative chemical exchange saturation transfer (CEST) imaging with magnetic resonance fingerprinting (MRF),” Magn. Reson. Med., vol. 80, no. 6, pp. 2449–2463, 2018, doi: 10.1002/mrm.27221.

[6] O. Cohen et al., “CEST MR fingerprinting (CEST-MRF) for brain tumor quantification using EPI readout and deep learning reconstruction,” Magn. Reson. Med., vol. 89, no. 1, pp. 233–249, 2023, doi: 10.1002/mrm.29448.

[7] O. Perlman, K. Herz, M. Zaiss, O. Cohen, M. S. Rosen, and C. T. Farrar, “CEST MR-Fingerprinting: practical considerations and insights for acquisition schedule design and improved reconstruction,” Magn. Reson. Med., vol. 83, no. 2, pp. 462–478, 2020, doi: 10.1002/mrm.27937.

[8] O. Perlman et al., “Quantitative imaging of apoptosis following oncolytic virotherapy by magnetic resonance fingerprinting aided by deep learning,” Nat. Biomed. Eng., vol. 6, no. 5, pp. 648–657, 2022, doi: 10.1038/s41551-021-00809-7.

[9] O. Cohen and R. Otazo, “Global deep learning optimization of chemical exchange saturation transfer magnetic resonance fingerprinting acquisition schedule,” NMR Biomed., vol. 36, no. 10, p. e4954, 2023, doi: 10.1002/nbm.4954.

[10] O. Cohen, R. J. Young, and R. Otazo, “Quantitative multislice and jointly optimized rapid CEST for in vivo whole-brain imaging,” Magn. Reson. Med., vol. 94, no. 2, pp. 541–553, 2025, doi: 10.1002/mrm.30488.

[11] B. Kim, M. Schär, H. Park, and H.-Y. Heo, “A deep learning approach for magnetization transfer contrast MR fingerprinting and chemical exchange saturation transfer imaging,” Neuroimage, vol. 221, p. 117165, 2020, doi: 10.1016/j.neuroimage.2020.117165.

[12] N. Vladimirov, O. Cohen, H.-Y. Heo, M. Zaiss, C. T. Farrar, and O. Perlman, “Quantitative molecular imaging using deep magnetic resonance fingerprinting,” Nat. Protoc., vol. 20, no. 10, pp. 3024–3054, 2025, doi: 10.1038/s41596-025-01152-w.

[13] L. Feng, “Golden-Angle Radial MRI: Basics, Advances, and Applications,” J. Magn. Reson. Imaging, vol. 56, no. 1, pp. 45–62, 2022, doi: 10.1002/jmri.28187.

[14] O. Cohen, E. J. Sutton, R. J. Young, and R. Otazo, “Quantitative chemical exchange saturation transfer imaging with golden-angle radial k-space and locally low-rank reconstruction,” NMR Biomed., 2026, doi: 10.1002/nbm.70428.

[15] L. Feng et al., “Golden-angle radial sparse parallel MRI: combination of compressed sensing, parallel imaging, and golden-angle radial sampling for fast and flexible dynamic volumetric MRI,” Magn. Reson. Med., vol. 72, no. 3, pp. 707–717, 2014, doi: 10.1002/mrm.24980.

[16] V. Murray, S. Siddiq, C. Crane, M. El Homsi, T.-H. Kim, C. Wu, and R. Otazo, “Movienet: Deep space-time-coil reconstruction network without k-space data consistency for fast motion-resolved 4D MRI,” Magn. Reson. Med., vol. 91, no. 2, pp. 600–614, 2024, doi: 10.1002/mrm.29892.

[17] M. Buehrer, K. P. Pruessmann, P. Boesiger, and S. Kozerke, “Array compression for MRI with large coil arrays,” Magn. Reson. Med., vol. 57, no. 6, pp. 1131–1139, 2007, doi: 10.1002/mrm.21237.

[18] J. A. Fessler, “On NUFFT-based gridding for non-Cartesian MRI,” J. Magn. Reson., vol. 188, no. 2, pp. 191–195, 2007, doi: 10.1016/j.jmr.2007.06.012.

[19] D. O. Walsh, A. F. Gmitro, and M. W. Marcellin, “Adaptive reconstruction of phased array MR imagery,” Magn. Reson. Med., vol. 43, no. 5, pp. 682–690, 2000, doi: 10.1002/(SICI)1522-2594(200005)43:5<682::AID-MRM10>3.0.CO;2-G.

[20] B. Recht, M. Fazel, and P. A. Parrilo, “Guaranteed minimum-rank solutions of linear matrix equations via nuclear norm minimization,” SIAM Rev., vol. 52, no. 3, pp. 471–501, 2010, doi: 10.1137/070697835.

[21] O. Cohen, B. Zhu, and M. S. Rosen, “MR fingerprinting deep reconstruction network (DRONE),” Magn. Reson. Med., vol. 80, no. 3, pp. 885–894, 2018, doi: 10.1002/mrm.27198.

[22] D. P. Kingma and J. Ba, “Adam: a method for stochastic optimization,” in Proceedings of International Conference on Learning Representations, San Diego, CA, USA, May 2015.

[23] N. Otsu, “A threshold selection method from gray-level histograms,” IEEE Trans. Syst. Man Cybern., vol. 9, no. 1, pp. 62–66, 1979, doi: 10.1109/TSMC.1979.4310076.

[24] Z. Wang, A. C. Bovik, H. R. Sheikh, and E. P. Simoncelli, “Image quality assessment: From error visibility to structural similarity,” IEEE Trans. Image Process., vol. 13, no. 4, pp. 600–612, 2004, doi: 10.1109/TIP.2003.819861.

[25] R. G. Pontius Jr., O. Thontteh, and H. Chen, “Components of information for multiple resolution comparison between maps that share a real variable,” Environmental and Ecological Statistics, vol. 15, no. 2, pp. 111–142, 2008, doi: 10.1007/s10651-007-0043-y.

[26] V. Murray, Y. Wen, S. Erattakulangara, O. Akin, R. Do, G. Behr, Z. Zhang, A. Guidon, and R. Otazo, “3D and 4D Free-Breathing Abdominal T1-Weighted MRI in Clinical Practice Using Deep Learning Auto-Navigation and Reconstruction,” Magn. Reson. Med., vol. 96, no. 5, pp. 2266–2277, 2026, doi: 10.1002/mrm.70505.

[27] J. Jing, A. Mekhanik, M. Schellenberg, V. Murray, O. Cohen, and R. Otazo, “Combination of deep learning reconstruction and quantification for dynamic contrast-enhanced (DCE) MRI,” Magn. Reson. Imaging, vol. 117, p. 110310, 2025, doi: 10.1016/j.mri.2024.110310.

[28] Y. Wang, J. Wen, J. Xu, et al., “Accelerated chemical exchange saturation transfer imaging with deep unrolling networks and synthetic brain tumor datasets,” Magn. Reson. Med., vol. 95, no. 4, pp. 2009–2022, 2026, doi: 10.1002/mrm.70172.

[29] J. Xu, T. Zu, Y.-C. Hsu, X. Wang, K. W. Y. Chan, and Y. Zhang, “Accelerating CEST imaging using a model-based deep neural network with synthetic training data,” Magn. Reson. Med., vol. 91, no. 2, pp. 583–599, 2024, doi: 10.1002/mrm.29889.

## Figure legends

Figure 1. Movienet architecture and supervised reconstruction pathway. For each slice, undersampled magnitude images from 8 virtual coils and 30 measurements are flattened to 240 channels and processed by a residual encoder-decoder. The output is a 30-frame image series trained against the matched sharpened GRASP-LLR target with Smooth L1 loss.

Figure 2. Image-domain performance versus radial spoke count in the primary held-out retrospective examination. Points and lines show the median across 900 matched slice-frame pairs (30 slices × 30 frames), and the light-blue band shows the interquartile range. NRMSE is displayed as a percentage. The vertical dashed line marks the 8-spoke operating point selected for prospective acquisition. Arrows in the panel titles indicate whether increasing or decreasing values represent improved performance for each metric.

Figure 3. Retrospective image-domain comparison at frame 20 in representative inferior, middle, and superior axial slices from the held-out healthy test volunteer. The 34-spoke GRASP-LLR reference images are compared with 8-spoke GRASP-LLR, Movienet-8, and Movienet-34 reconstructions; difference images are shown for both Movienet reconstructions.

Figure 4. Retrospective quantitative CEST map-domain comparison in the held-out healthy test volunteer. Reference maps fitted from original, unsharpened 34-spoke GRASP-LLR images are compared with Movienet-8 and Movienet-34 maps for T1w, T2w, ksw, kssw, fs, and fss.

Figure 5. Evaluation of the prospectively acquired 8-spoke data in 3 representative slices of a healthy subject. Rows show representative inferior, middle, and superior axial slices for one frame of the MRF schedule. From left to right, columns show the GRASP-LLR reference reconstruction with 34 spokes, retrospective Movienet-8 reconstruction, retrospective difference image, prospective Movienet-8 reconstruction, and prospective difference image.

Figure 6**.** Quantitative CEST-MRF maps from a prospective 8-spoke acquisition in a representative slice. From left to right, columns show the reference map fitted from the original paired 34-spoke GRASP-LLR acquisition, retrospective Movienet-8 map, retrospective difference map, prospective Movienet-8 map, and prospective difference map. Rows show T1w, T2w, ksw, kssw, fs, and fss. Parameter and difference-map units and color scales are displayed beside the panels.

Figure 7. Healthy 64-spoke high-sampling-density evaluation. Image-domain comparison at frame 20 across three representative axial slices, using the 64-spoke GRASP-LLR reconstruction as the reference.

Figure 8. Quantitative CEST map-domain comparison in a representative slice; the reference maps were fitted from the 64-spoke GRASP-LLR reconstruction and compared with Movienet-8 and Movienet-34 maps.

**Table 1. Image reconstruction metrics in healthy-subject evaluations**

| Evaluation | Reconstruction | SSIM | PSNR (dB) | MAE | NRMSE (%) | Pearson correlation |
|---|---|---|---|---|---|---|
| **Retrospective** | **Movienet-8** | 0.842 ± 0.061 | 24.7 ± 2.1 | 0.016 ± 0.006 | 7.1 ± 1.8 | 0.953 ± 0.026 |
| **Retrospective** | **Movienet -34** | 0.913 ± 0.037 | 27.9 ± 1.8 | 0.010 ± 0.004 | 4.6 ± 1.2 | 0.982 ± 0.011 |
| **Prospective** | **Movienet -8** | 0.831 ± 0.064 | 24.2 ± 2.2 | 0.017 ± 0.006 | 7.4 ± 1.9 | 0.948 ± 0.028 |
| **64-spoke reference** | **Movienet -8** | 0.821 ± 0.058 | 23.9 ± 2.0 | 0.018 ± 0.006 | 7.6 ± 1.7 | 0.962 ± 0.014 |
| **64-spoke reference** | **Movienet -34** | 0.887 ± 0.044 | 26.8 ± 1.7 | 0.012 ± 0.004 | 4.8 ± 1.1 | 0.987 ± 0.006 |

Values are mean ± standard deviation across 30 slices and 30 contrast frames per acquisition. NRMSE is 100 × RMSE divided by the reference 99th–1st percentile range within each slice and frame. References are original 34-spoke GRASP-LLR for the retrospective rows, same-acquisition 8-spoke GRASP-LLR for the prospective row, and 64-spoke GRASP-LLR for the final two rows. Higher SSIM, PSNR, and Pearson correlation and lower MAE and NRMSE indicate better agreement.

**Table 2. Tissue map metrics in healthy subject evaluations**

| Evaluation | Reconstruction | SSIM | PSNR (dB) | MAE | NRMSE (%) | Pearson correlation |
|---|---|---|---|---|---|---|
| **Retrospective** | **Movienet -8** | 0.786 ± 0.084 | 22.1 ± 3.5 | 0.094 ± 0.041 | 18.2 ± 7.1 | 0.821 ± 0.183 |
| **Retrospective** | **Movienet -34** | 0.852 ± 0.069 | 24.8 ± 3.1 | 0.066 ± 0.032 | 13.2 ± 5.4 | 0.892 ± 0.146 |
| **Prospective** | **Movienet -8** | 0.771 ± 0.088 | 21.7 ± 3.6 | 0.099 ± 0.044 | 19.0 ± 7.5 | 0.814 ± 0.194 |
| **Prospective** | **Movienet -34** | 0.838 ± 0.073 | 24.2 ± 3.2 | 0.071 ± 0.034 | 14.1 ± 5.8 | 0.871 ± 0.154 |
| **64-spoke reference** | **Movienet -8** | 0.741 ± 0.096 | 20.8 ± 3.7 | 0.111 ± 0.052 | 21.4 ± 8.6 | 0.832 ± 0.221 |
| **64-spoke reference** | **Movienet -34** | 0.813 ± 0.081 | 23.6 ± 3.4 | 0.078 ± 0.039 | 16.1 ± 6.4 | 0.864 ± 0.174 |

Values are mean ± standard deviation across 30 slices and six fitted maps per acquisition. NRMSE uses the reference 99th–1st percentile range. Metrics use a common valid-voxel mask. References are original 34-spoke GRASP-LLR, except for the final two rows, which use 64-spoke GRASP-LLR. Higher SSIM, PSNR, and Pearson correlation and lower MAE and NRMSE indicate better agreement.

**Supporting Table S1. Subject-level image-domain leave-one-subject-out evaluation of Movienet-8**

| Volunteer | Acquisitions | Pearson correlation | NRMSE (%) |
|---|---|---|---|
| 1 | 2 | 0.977 | 6.9 |
| 2 | 2 | 0.971 | 6.5 |
| 3 | 2 | 0.974 | 6.9 |
| 4 | 2 | 0.962 | 7.5 |
| 5 | 2 | 0.949 | 11.7 |
| 6 | 2 | 0.958 | 9.2 |
| 7 | 2 | 0.958 | 7.5 |
| 8 | 2 | 0.968 | 6.5 |
| 9 | 2 | 0.975 | 7.5 |
| 10 | 2 | 0.938 | 8.9 |

Each row represents a different held-out volunteer. Pearson correlation and NRMSE were calculated within the reference-derived mask separately for each slice and frame, as in Table 1. NRMSE is 100 × RMSE / (P99 − P1), where P99 and P1 are the 99th and 1st percentiles of the masked reference intensities for that slice and frame. The 900 slice–frame values were averaged within each acquisition, then the two acquisition means were averaged for each volunteer. Median subject-level correlation was 0.965 and median NRMSE was 7.5%.

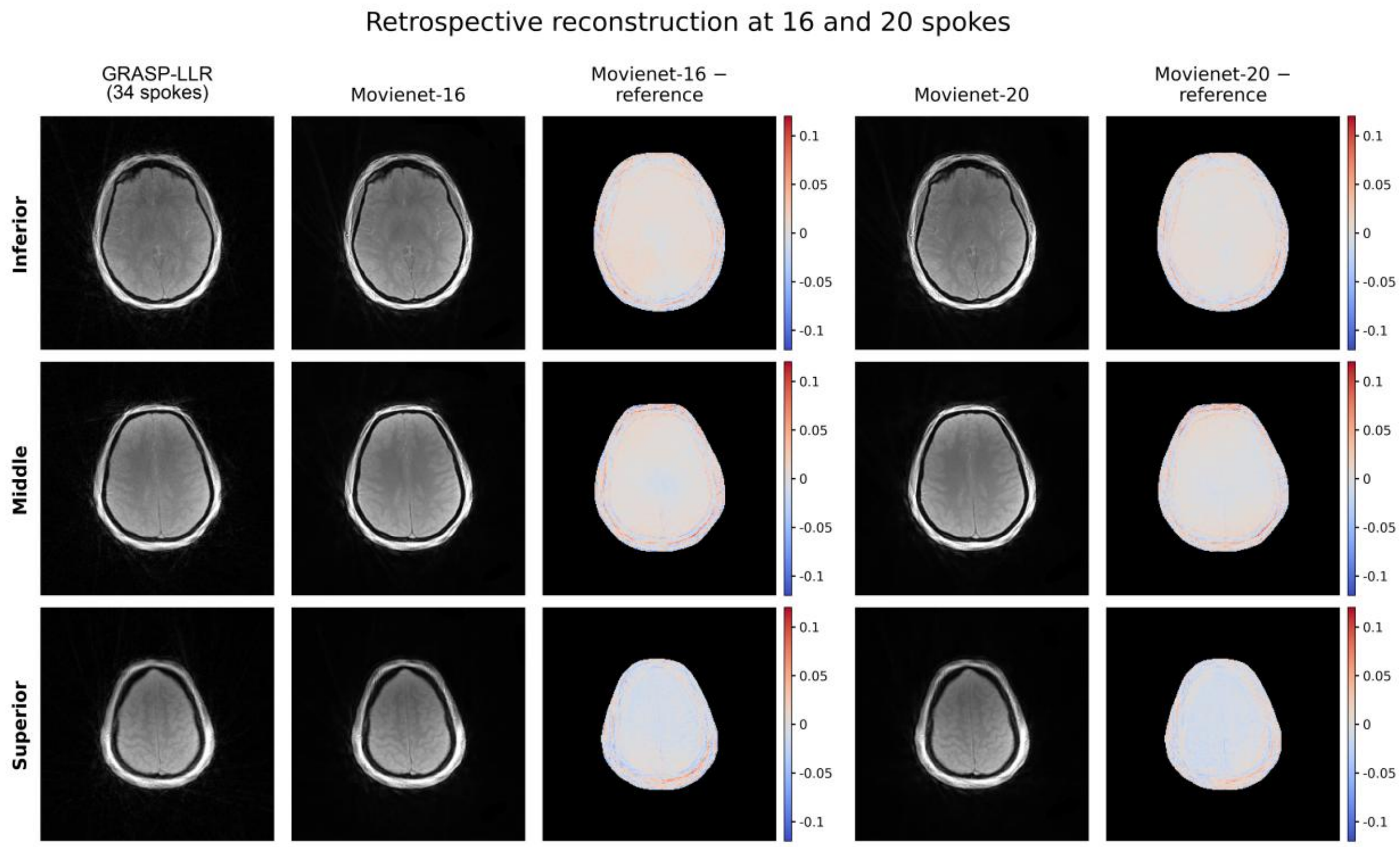

**Supporting Figure S1.** Retrospective image reconstruction using 16 and 20 spokes per measurement in the primary held-out healthy examination. Rows show representative inferior, middle, and superior axial slices (15, 21, and 25, respectively) at measurement 20. Columns show the 34-spoke GRASP-LLR reference, Movienet-16 reconstruction, Movienet-16 minus reference, Movienet-20 reconstruction, and Movienet-20 minus reference. A common grayscale window is applied to all reconstructions within each row. Signed difference images are displayed within the same reference-derived brain mask using a fixed range of −0.12 to 0.12 in normalized image-intensity units.

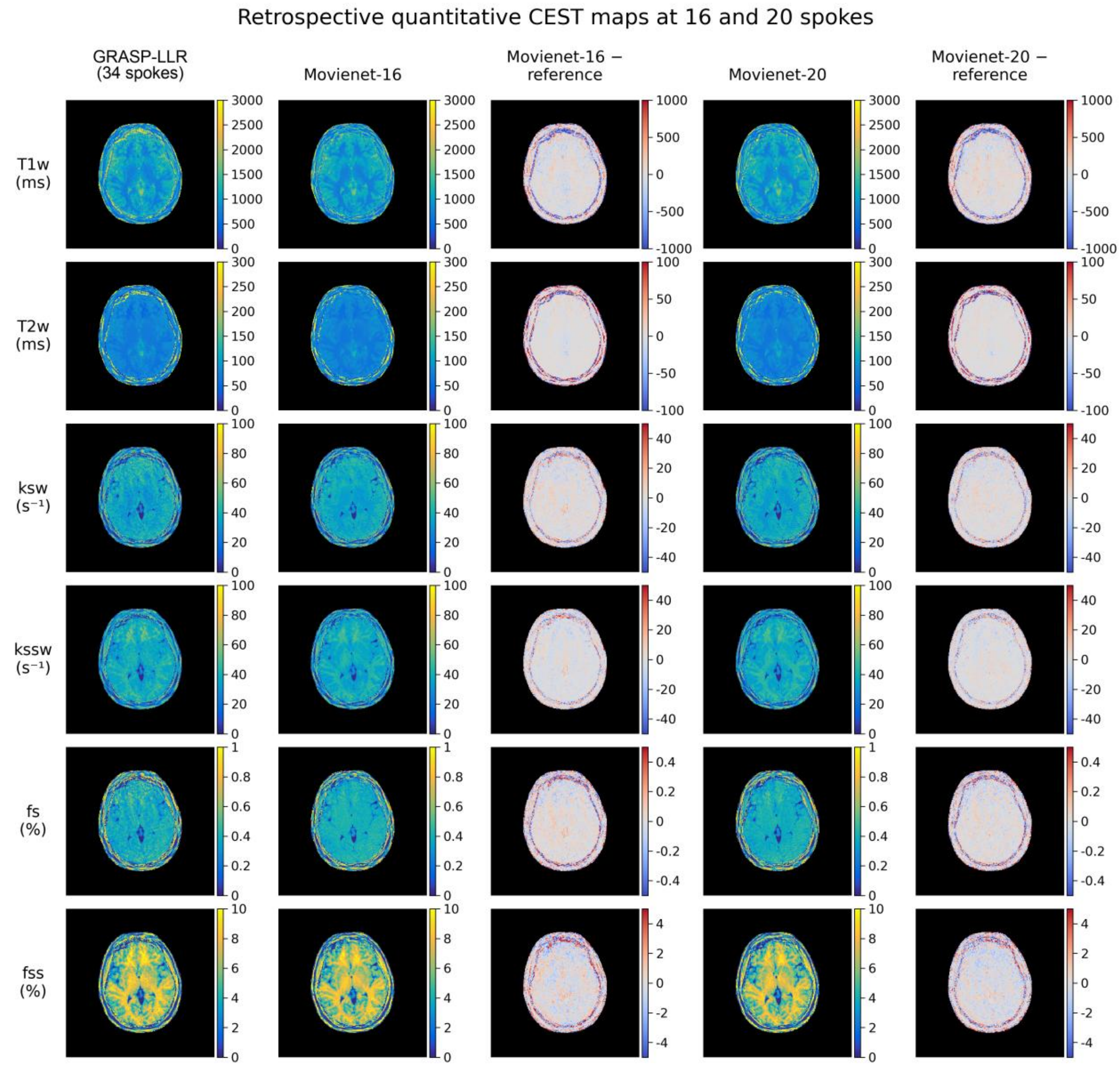

**Supporting Figure S2.** Quantitative CEST-MRF maps from retrospective Movienet-16 and Movienet-20 reconstruction in the inferior axial slice (slice 15) shown in Figure S1. Reference maps were fitted from the original, unsharpened 34-spoke GRASP-LLR images. Columns show the reference, Movienet-16, Movienet-16 minus reference, Movienet-20, and Movienet-20 minus reference. Rows show T1w, T2w, ksw, kssw, fs, and fss. Maps were estimated from all 30 measurements using the same DRONE model. Fractions are displayed as percentages.

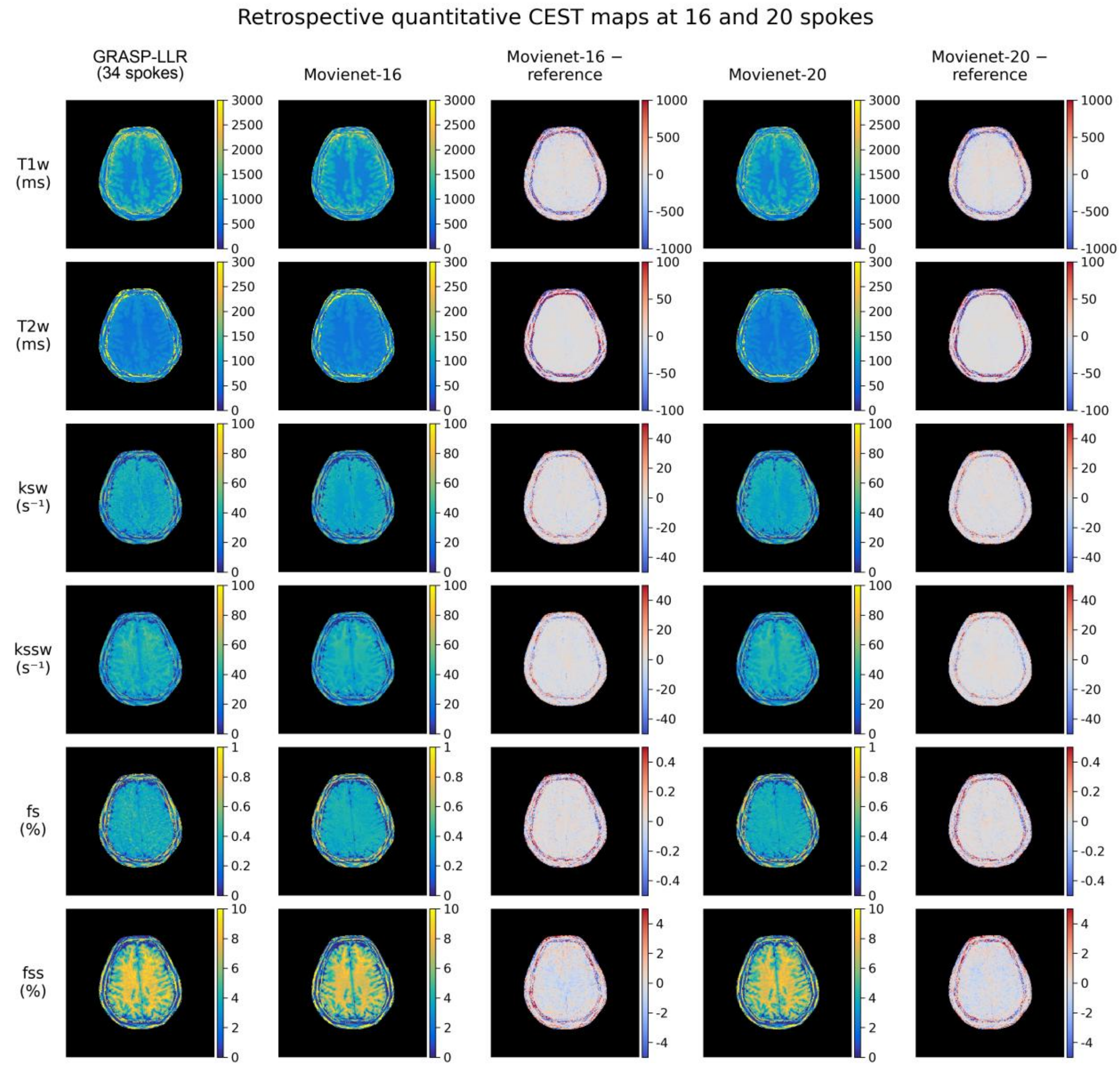

**Supporting Figure S3.** Quantitative CEST-MRF maps from retrospective Movienet-16 and Movienet-20 reconstruction in the middle axial slice (slice 21) shown in Figure S1. Reference maps were fitted from the original, unsharpened 34-spoke GRASP-LLR images. Columns show the reference, Movienet-16, Movienet-16 minus reference, Movienet-20, and Movienet-20 minus reference. Rows show T1w, T2w, ksw, kssw, fs, and fss. Maps were estimated from all 30 measurements using the same DRONE model. Fractions are displayed as percentages.

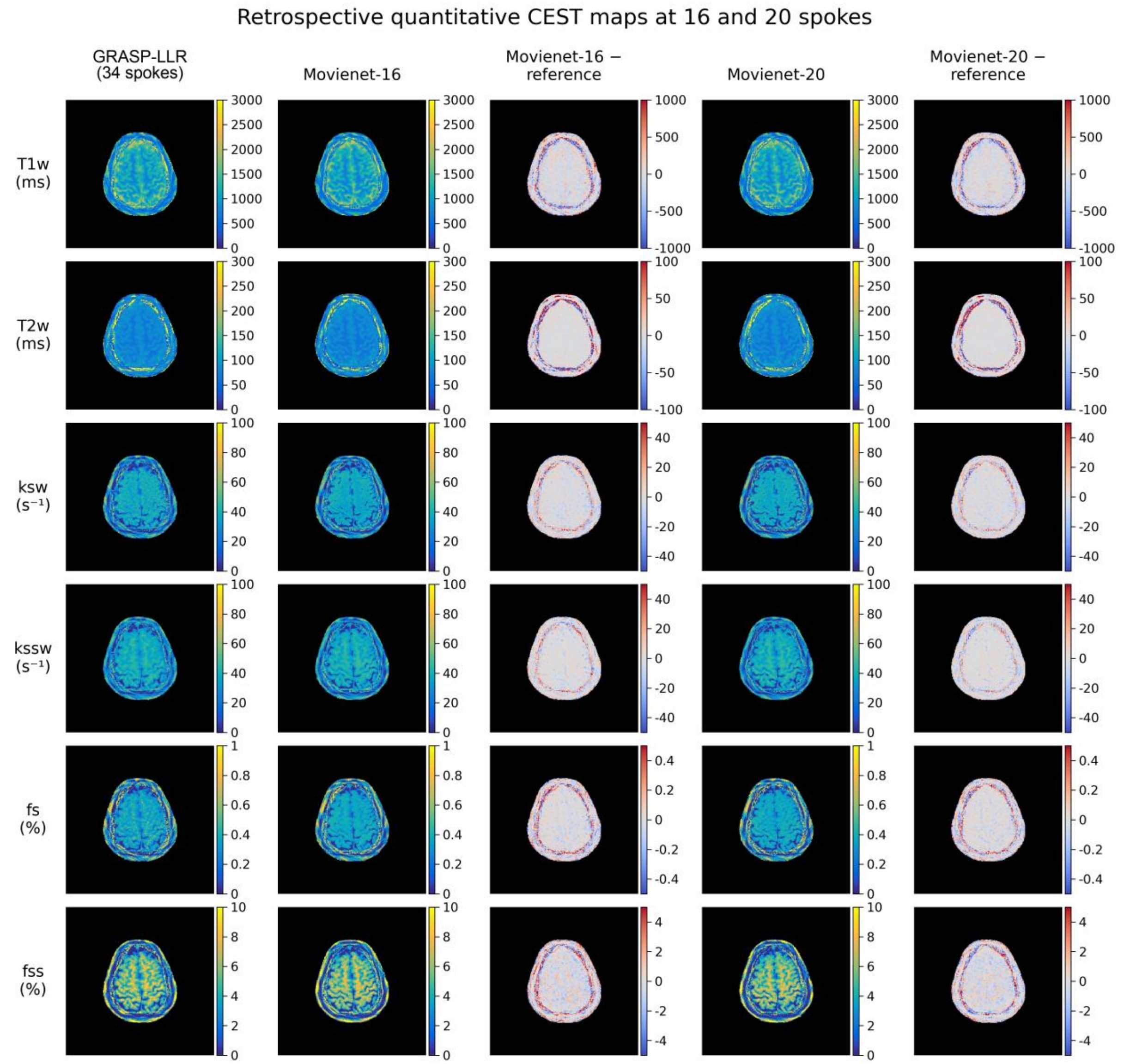

**Supporting Figure S4.** Quantitative CEST-MRF maps from retrospective Movienet-16 and Movienet-20 reconstruction in the superior axial slice (slice 25) shown in Figure S1. Reference maps were fitted from the original, unsharpened 34-spoke GRASP-LLR images. Columns show the reference, Movienet-16, Movienet-16 minus reference, Movienet-20, and Movienet-20 minus reference. Rows show T1w, T2w, ksw, kssw, fs, and fss. Maps were estimated from all 30 measurements using the same DRONE model. Fractions are displayed as percentages.